\documentstyle[aps,amssymb,preprint,epsf,tighten]{revtex}
\begin{document}
\draft

\title{The incorporation of matter into characteristic numerical relativity}

\author{Nigel T. Bishop${}^{1}$,
        Roberto G\'omez${}^{2}$,
	Luis Lehner${}^{2,3}$,
        Manoj Maharaj${}^{4}$ and
        Jeffrey Winicour${}^{2}$}
\address{
${}^{1}$Department of Mathematics, Applied Mathematics and Astronomy,\\
University of South Africa, P.O. Box 392, Pretoria 0003, South Africa \\
${}^{2}$Department of Physics and Astronomy,\\
University of Pittsburgh, Pittsburgh, PA 15260\\
${}^{3}$Center for Relativity, \\
The University of Texas at Austin, Austin TX 78712 \\
${}^{4}$Department of Mathematics and Applied Mathematics,\\
University of Durban-Westville, Durban 4000, South Africa \\
 }

\maketitle

\begin{abstract}
A code that implements Einstein equations in the characteristic formulation
in $3D$ has been developed and thoroughly tested for the vacuum case. Here,
we describe how to incorporate matter, in  the form of a perfect fluid,
into the
code. The extended code has been written and validated in a number of
cases. It is stable and capable of contributing towards an 
understanding of a number of problems in black hole astrophysics.


\end{abstract}

\section{Introduction}

A code based on the characteristic formulation of numerical relativity has been 
developed for the general 3D problem~\cite{hpn}. The code computes the 
gravitational field of the full Einstein equations for a vacuum spacetime 
between some inner timelike worldtube $\Gamma$ and future null infinity. The 
code has been tested with the inner worldtube $\Gamma$ being the (past) event 
horizon of a Schwarzschild black hole, with incoming gravitational radiation 
scattered off the black hole. The tests have included both the linear and 
highly nonlinear regimes, and show the robustness of the characteristic 
formulation in the study of radiative problems. Here we consider whether 
there are real astrophysical problems to which the characteristic code alone 
could be applied.

Astrophysical problems that can be tackled by the code require the existence of 
a natural inner worldtube $\Gamma$ on which boundary data is known. This would 
include the (past) event horizon of a black hole. In principle the black hole 
could be of Kerr-Newman type, but at present suitable boundary and initial data
(for the characteristic formulation) are known only in the Schwarzschild case, 
and also in some multi-hole vacuum spacetimes~\cite{ll98}. However, if matter is 
incorporated 
into the code, then the range of possible astrophysical applications can be 
greatly extended. It would then in principle be possible to compute, in full 
nonlinear general relativity, the gravitational field and matter flow of a black 
hole accreting dust and gas, and of a black hole capturing polytropic or more
realistic approximations to 
neutron stars. This paper does not solve any of these important astrophysical 
problems, but rather it lays the groundwork for doing so.

The incorporation of matter into a characteristic code has been discussed 
previously. An axisymmetric code with matter has been used to consider a problem 
in cosmology~\cite{ntb96}; and a spherically symmetric matter code, that 
includes Cauchy-characteristic matching, has been reported 
recently~\cite{dubal98,mat1D}. In contrast, the present work has no symmetry 
requirement.

The paper restricts attention to matter in the form of a perfect fluid, by which 
we mean that the pressure $p$ is a function {\em only} of the density $\rho$. In 
section \ref{s:form} we summarize known results, and introduce our notation, for 
the characteristic formalism and perfect fluid evolution. Next, in section 
\ref{s:ntheory} we derive the fluid evolution equations in our characteristic 
coordinates; we also find the additional terms (compared to the vacuum case) 
that appear in the Einstein equations. Section \ref{s:num} describes details of 
the numerical implementation of the fluid evolution equations (the additional 
terms in the Einstein equations are straightforward to code). The resulting 
matter plus gravity code has been run on a variety of test problems, and the 
results are described in section \ref{s:test}. In all cases, the code is found 
to be stable and convergent. We end with a Conclusion (section \ref{s:conc}), 
and an Appendix that contains expressions from section \ref{s:ntheory} that are 
rather long.

\section{Formalism}
\label{s:form}

\subsection{The null cone - previous results}
\label{ncold}

The formalism for the numerical evolution of Einstein's equations, in null cone 
coordinates, is well known~\cite{hpn,cce} (see 
also~\cite{ntb93,ntb90,rai83,bondi}). Nevertheless, for the sake of 
completeness, 
we give here a summary of the formalism, including the necessary equations. We 
will use the notation and language of~\cite{hpn}.

We use coordinates based upon a family of outgoing null hypersurfaces.
We let $u$ label these hypersurfaces, $x^A$ $(A=2,3)$, label
the null rays and $r$ be a surface area coordinate. In the resulting
$x^\alpha=(u,r,x^A)$ coordinates, the metric takes the Bondi-Sachs
form~\cite{bondi,sachs}
\begin{eqnarray}
   ds^2 & = & -\left(e^{2\beta}(1 + {W \over r}) -r^2h_{AB}U^AU^B\right)du^2
        -2e^{2\beta}dudr -2r^2 h_{AB}U^Bdudx^A \nonumber \\
        & + & r^2h_{AB}dx^Adx^B,    \label{eq:bmet}
\end{eqnarray}
where $W$ is related to the more usual Bondi-Sachs variable $V$ by
$V=r+W$; and where $h^{AB}h_{BC}=\delta^A_C$ and
$det(h_{AB})=det(q_{AB})$, with $q_{AB}$ a unit sphere metric.  In
analyzing the Einstein equations, we also use the intermediate variable
\begin{equation}
Q_A = r^2 e^{-2\,\beta} h_{AB} U^B_{,r}.
\end{equation}

We work in stereographic coordinates $x^A=(q,p)$ for which the unit sphere
metric is
\begin{equation}
q_{AB} dx^A dx^B = \frac{4}{P^2}(dq^2+dp^2),
\end{equation}
where
\begin{equation}
        P=1+q^2+p^2.
\end{equation}
We also introduce a complex dyad $q_A$ defined by
\begin{equation}
      q^A=\frac{P}{2}(1,i)
\end{equation}
with $i=\sqrt{-1}$. For an arbitrary Bondi-Sachs metric,
$h_{AB}$ can then be represented by its dyad component
\begin{equation}
J=h_{AB}q^Aq^B/2,
\end{equation}
with the spherically symmetric case characterized by $J=0$. The
full nonlinear $h_{AB}$ is uniquely determined by $J$, since the
determinant condition implies that the remaining dyad component
\begin{equation}
K=h_{AB}q^A \bar q^B /2
\end{equation}
satisfies $1=K^2-J\bar J$.  We also introduce
spin-weighted fields 
\begin{equation}
U=U^Aq_A, \; \; \; Q=Q_Aq^A,
\end{equation}
 as well as the (complex differential) eth operators $\eth$ and $\bar \eth$
 (see~\cite{eth} for full details).

The Einstein equations $G_{ab}=0$ decompose into hypersurface equations, 
evolution equations and conservation laws. Naturally, the  
equations will require additional terms to allow for the presence of matter, and 
the extended equations are given later in section \ref{sec:matein}. Here we just 
note that the hypersurface equations form a hierarchical set for $\beta_{,r}$, 
$(r^2 Q)_{,r}$, $U_{,r}$ and $W_{,r}$; and the evolution equation is an 
expression for $(rJ)_{,ur}$.

The remaining independent equations are the conservation conditions, but they 
will not be needed here.

The null cone problem is normally formulated in the region of spacetime between  
a timelike or null worldtube $\Gamma$ and ${\cal I}^+$, with initial data $J$ given on 
the null cone $u=0$ in this domain. Boundary data for $\beta$, $Q$, $U$, $W$ and 
$J$ is also required on $\Gamma$. The metric variables used remain regular at 
${\cal I}^+$, and we represent ${\cal I}^+$ on a finite grid by using a 
compactified radial coordinate $x=r/(1+r)$.

\subsection{Perfect fluid - previous results}

The description of a perfect fluid is well understood (e.g.~\cite{HE,Steph}).
The stress-energy tensor is
\begin{equation}
T_{ab}=(\rho + p) v_a v_b + p g_{ab}
\end{equation}
where $\rho$ is the density, $p$ is the pressure, $v_a$ is the 4-velocity
and $g_{ab}$ is the metric. In the cases considered here, the pressure depends 
{\em only} on the density, i.e.
\begin{equation}
p=p(\rho).
\end{equation}
The matter evolution equations follow from the conservation law 
$T^{ab}{}_{;b}=0$, and are
\begin{equation}
\rho_{,a}v^a + (\rho+p)v^a_{;a}=0,\; \;
(\rho+p)v_{a;b}v^b + (\delta^b_a + v_a v^b)p_{,b}=0,
\end{equation}
which may be rewritten as
\begin{eqnarray}
\rho_{,a} v_b g^{ab} + (\rho + p) (v_{a,b} - \Gamma^c_{ab} v_c) g^{ab}=0
\label{eq:mev0} \\
M_a \equiv (\rho + p) (v_{a,b} - \Gamma^c_{ab} v_c) v_d g^{bd} + p_{,a}
+ v_a v_c p_{,b} g^{bc} =0.  \label{eq:meva}
\end{eqnarray}
The numerical implementation of the fluid equations coupled to G.R. has been
investigated primarily within the $3+1$ or Cauchy framework.
In this formulation, the spacetime is foliated by a sequence of spacelike 
surfaces,
initial data is given on an initial surface  and the evolution equations are 
used
to compute the future. The numerical investigation of the ``G.R.-Hydro'' problem
started in the early 70's by Wilson~\cite{wilson} and since then it has received
considerable attention. The difficulty of modeling these equations has spurred
the development of sophisticated techniques to deal with the diverse 
idiosyncracies of the problem. Thus, artificial viscosity techniques, total 
variation diminishing flux limiters, shock-capturing schemes, etc. are
actively employed to aid in the numerical modeling; refer 
to~\cite{grhydro} for a recent review.

Another  problem one faces when attempting a numerical simulation in the $3+1$ 
formalism is
that an artificial ``outer boundary'' has to be included at some radius in order 
to
deal with a finite grid. This introduces spurious reflections that spoil long 
term
evolutions.  In the characteristic formulation, on the other hand, one can use
Penrose's compactification techniques to include infinity in the numerical grid. 
Additionally,
being able to access null infinity allows one to obtain physical quantities, 
like
the radiation given off by the system and the total mass, unambiguously. 
There is not as much experience with the G.R.-Hydro problem  in the 
characteristic
formulation, but the results obtained so far are encouraging, indicating
the characteristic formulation of G.R. might be a valuable tool to study a 
variety
of astrophysical problems.

\section{Perfect fluid in the null cone formalism}
\label{s:ntheory}

\subsection{Perfect fluid equations}

In order to proceed further we represent the angular part of the 4-velocity by 
means of the complex quantity
\begin{equation}
V_{ang}=q^A v_A=\frac{P}{2}(v_2 + i v_3),
\end{equation}
where the ${}_{ang}$ suffix is introduced to avoid confusion with the
Bondi-Sachs metric variable $V$. The matter evolution equations are then:
\begin{itemize}
\item equation (\ref{eq:mev0});
\item equation (\ref{eq:meva}) in the form
\begin{equation}
M_1=0, \; \; \; q^A M_A=0.
\label{eq:Mev}
\end{equation}
\end{itemize}
We also introduce the notation
\begin{equation}
p_\rho=\frac{1}{p+\rho} \frac{dp}{d\rho}
\label{eq:prho}
\end{equation}
and then write
\begin{equation}
p_{,a}=p_\rho (p+\rho) \rho_{,a}.
\end{equation}
The result of doing this is that, in the matter evolution equations, there is no 
explicit division by $(p+\rho)$ (which could be zero). From a numerical point of 
view, it is possible to write the procedure for computing $p_\rho$ so as to 
ensure its good behavior in the low density limit.

The matter evolution equations have been calculated using Maple; they are
rather long and are given in the Appendix. The denominators are important in
determining whether there may be singular behavior, and here we note that
the form of the equations is
\begin{equation}
\rho_{,u}=\frac{1}{r^3 (v_1)^2 (\frac{dp}{d\rho} -1)} F_\rho
\label{eq:rhou}
\end{equation}
\begin{equation}
v_{1,u}=\frac{1}{r^3 v_1 (\frac{dp}{d\rho} -1)} F_1
\label{eq:v1u}
\end{equation}
\begin{equation}
V_{ang,u}=\frac{1}{r^2 v_1} F_V.
\label{eq:Vu}
\end{equation}
Formally, the equations (\ref{eq:rhou}) and (\ref{eq:v1u}) could be singular if
$\frac{dp}{d\rho}$ were to be 1; but this would 
correspond physically to the velocity of sound being equal to that of light.
We do not concerns ourselves with those cases at the moment and defer its
treatment for a later work.
Also, the equations (\ref{eq:rhou}), (\ref{eq:v1u}) and (\ref{eq:Vu}) could be 
singular if  $v_1$ were to be zero; yet, starting from $-1=v_a v_b g^{ab}$,
\begin{equation}
v_1 e^{-2\beta} (\frac{V}{r} v_1 - 2 v_0 -2 U^A v_A) < -1,
\end{equation}
which can never be satisfied if $v_1=0$. Thus, from an analytic point of view, 
the equations (\ref{eq:rhou}), (\ref{eq:v1u}) and (\ref{eq:Vu}) form a 
well-behaved set of evolution equations.

Finally, again using the condition $-1=g^{ab} v_a v_b$, we obtain an expression 
for the remaining velocity component $v_0$
\begin{equation}
v_0=\frac{e^{2\beta}(2K V_{ang} \bar{V}_{ang}
- J \bar{V}^2_{ang} -\bar{J} V^2_{ang} + 2r^2) +2rv_1(V v_1
- r U \bar{V}_{ang} - r \bar{U} V_{ang})}{4 v_1 r^2}.
\label{eq:v0}
\end{equation}

\subsection{The Einstein equations}
\label{sec:matein}
The introduction of matter also amends the Einstein equations. We write the
equations as
\begin{equation}
R_{ab}=8 \pi (T_{ab} -\frac{1}{2} g_{ab} T),
\end{equation}
and note that
\begin{equation}
T=-(\rho + p) + 4 p =3p - \rho.
\end{equation}
Thus the Einstein equations for a perfect fluid are
\begin{eqnarray}
R_{ab}=8 \pi  (\rho + p) v_a v_b + g_{ab} (p-\frac{3p-\rho}{2}))
\nonumber \\
=8 \pi ((\rho + p) v_a v_b +g_{ab}\frac{\rho-p}{2}).
\label{eq:Rab}
\end{eqnarray}

In the expressions below, $N_\beta$, $N_U$, $N_Q$, $N_W$ and $N_J$ represent the 
nonlinear aspherical terms (in a sense specified in~\cite{cce}). These 
quantities are quite long and are not repeated here since they have been given 
explicitly in~\cite{hpn}. Using Maple we 
have confirmed that
\begin{itemize}
\item $R_{11}$ in equation (\ref{eq:Rab}) leads to
\begin{equation}
\beta_{,r}=N_\beta + 2 \pi r (\rho +p) (v_1)^2.
\label{eq:beta}
\end{equation}
\item $R_{1A} q^A$ in equation (\ref{eq:Rab}) leads to
\begin{equation}
(r^2 Q)_{,r}= -r^2 (\bar \eth J + \eth K)_{,r}
                +2r^4\eth \left(r^{-2}\beta\right)_{,r}
                 + N_Q +
                  16 \pi r^2 (\rho +p) v_1 V_{ang}.
\label{eq:Q}
\end{equation}
\item The equation for $U$ is a definition so the presence of matter does not 
change it:
\begin{equation}
U_{,r}=\frac{e^{2\beta}}{r^2}(K Q -J \bar{Q}).
\label{eq:U}
\end{equation}
\item $R_{AB} h^{AB}$ in equation (\ref{eq:Rab}) leads to
\begin{eqnarray}
W_{,r}= \frac{1}{2} e^{2\beta}{\cal R} -1
- e^{\beta} \eth \bar \eth e^{\beta}
+ \frac{1}{4} r^{-2} \left(r^4
                           \left(\eth \bar U +\bar \eth U \right)
                     \right)_{,r} 
+ N_W \nonumber \\
-4 \pi e^{2\beta}((\rho + p)
(K V_{ang} \bar{V}_{ang} -\frac{1}{2}(J \bar{V}^2_{ang}
+\bar{J} V^2_{ang})) + (\rho -p) r^2),
\label{eq:W}
\end{eqnarray}
where
\begin{equation}
{\cal R} =2 K - \eth \bar \eth K + \frac{1}{2}(\bar \eth^2 J + \eth^2 \bar J)
          +\frac{1}{4K}(\bar \eth \bar J \eth J - \bar \eth J \eth \bar J).
\end{equation}
\item $R_{AB} q^A q^B$ in equation (\ref{eq:Rab}) leads to
\begin{eqnarray}
    && 2 \left(rJ\right)_{,ur}
    - \left(r^{-1}V\left(rJ\right)_{,r}\right)_{,r} =
    -r^{-1} \left(r^2\eth U\right)_{,r}
    + 2 r^{-1} e^{\beta} \eth^2 e^{\beta}- \left(r^{-1} W \right)_{,r} J
    + N_J \nonumber \\
    && +\frac{4 e^{2\beta}\pi}{r}
    ( V^2_{ang} (\rho + p) +r^2 J (\rho -p)).
\label{eq:J}
\end{eqnarray}
\end{itemize}

\subsection{Summary}

The data required on the initial null cone is:
\begin{equation}
J \; \; \rho \; \; v_1 \; \; V_{ang}
\end{equation}
and these constitute the set of evolution variables. The auxiliary variables 
may then be determined on the initial null cone, and they are found in the 
following order: $p$ from the equation of state, $\beta$ from
equation (\ref{eq:beta}), $Q$ from equation (\ref{eq:Q}), $U$ from equation 
(\ref{eq:U}), $W$ from equation (\ref{eq:W}), and $v_0$ from  equation 
(\ref{eq:v0}).
The evolution equations (\ref{eq:J}), (\ref{eq:rhou}), (\ref{eq:v1u})
and (\ref{eq:Vu}) may now be used to find $J$, $\rho$, $v_1$ and $V_{ang}$
(in that order) on the ``next'' null cone.

In order to have a properly specified problem, we will also need boundary data 
on the timelike worldtube $\Gamma$. For the gravitational variables this data is 
$\beta$, $Q$, $U$, $W$ and $J$, and for the matter variables we will need 
$\rho$, $v_1$ and $V_{ang}$. 

\section{Numerical Implementation}
\label{s:num}
We constructed a set of algorithms to solve equations
(\ref{eq:beta}-\ref{eq:J}). In
discretizing the equations we follow closely the strategy developed for
the vacuum case~\cite{hpn}. We introduce a compactified radial
coordinate $x=r/(1+r)$ and define the numerical grid with coordinates
$(u^n, x_i, q_j, p_k) = (n \Delta u, 1/2 + (i-1) \Delta x, -1 + (j-3)
\Delta q, -1 + (j-3) \Delta p)$ (where $2 \Delta x=1/(N_x-1)$, and
$\Delta q = \Delta p = 2/(N_{\xi}-1)$ ) . Using finite differences to discretize
the equations, we center the derivatives at $({n+1/2}, i-1/2, j, k)$. 

The evolution equation is treated as in~\cite{hpn} where the right hand
side is modified to include the matter terms. Thus, the matter
variables at $({n+1/2}, i-1/2)$ are evaluated by taking the average
between the values at $(n+1, i-1)$ and $({n}, i)$, while the radial 
derivatives are obtained from the average of the corresponding values 
at $(n+1, i-3/2)$ and $({n}, i+1/2)$. 

Next, we proceed to integrate the evolution equation of the matter variables
by a simple iterative method which results in a second order in space, second 
order in time scheme, as follows.
First, note that the matter equations can be schematically written as
\begin{equation}
g_{,u} = F \, ,
\end{equation}
then, a direct second order discretization is obtained by
\begin{equation}
g^{n+1}_i = g^{n}_i + \Delta u F^{n+1/2}_{i} \, . \label{eq:iterat}
\end{equation}
Note that $F$ depends on the matter fields, thus without having at hand
the value of $g^{n+1}_i$, $F$ can not be directly evaluated at the midpoint.
However, the value of $F^{n+1/2}_{i}$ can indeed be obtained by a 
straightforward 
iteration.
In the first step we set $F^{n+1/2}_{i}=F^{n}_{i}$ and use it to
evaluate the right hand side of (\ref{eq:iterat}); then, the obtained 
approximation to $g^{n+1}_i$ is used to obtain a better approximation
to $F^{n+1/2}_i$ and so on. This
procedure is repeated a sufficient number of times to ensure second order
convergence of the algorithm.
However, care must be taken at the horizon where fields diverge which
will spoil the numerical modeling of the problem. If we had at hand a solution
that we could use to evaluate the fields near the horizon, we could just
integrate away from it. Unfortunately, this is not the case, and a
different strategy must be employed.

In this work, where we mainly investigate the suitability of the characteristic
formulation as a mean to simulate matter coupled to G.R.; we resort to evolving
the matter equations from the first point outside the worldtube to null
infinity. Hence, the previously described algorithm, where $F$ is evaluated at
the {\em i-th} point, can not be used since $F$ involves radial derivatives.
Hence, for the first point we use a different algorithm by employing a {\it one
sided} scheme taking backward differences to evaluate $F$. This scheme is only
first order accurate and its resulting discretization error does not match
smoothly to the one obtained with the second order scheme; thus, we do not
expect to obtain global second order accuracy in our numerical integration.
Several other alternatives are being investigated to obtain global second order
accuracy but we defer their implementation to a future work.

Finally, the hypersurface equations are discretized as in~\cite{hpn},
where the right hand sides now include the matter variables evaluated
at $(n+1, i-1/2)$ (which is straightforward having the matter fields
values at $(n+1,i)$ obtained in the previous step).

\section{Tests and Results}
\label{s:test}

We test the code by considering an initial localized distribution of matter 
around a Schwarzschild black hole with mass taken to be $M=1$. The gravitational 
initial data is taken as
\begin{equation}
J(u=0,r,x^A)=0.
\end{equation}
For a non-spherical initial distribution of matter, this condition is in a sense 
unphysical in that it will introduce spurious incoming gravitational 
radiation~\cite{rai85,pap98}; nevertheless it is simple and is suitable for code 
testing. The gravitational boundary data on the worldtube $\Gamma$, which we 
take to be the (past) event horizon of the black hole at $r=2$, is~\cite{hpn}
\begin{equation}
\beta=0, \; \; Q=0, \; \; U=0, \; \; W=-2, \; \; J=0.
\end{equation}
The initial data for the tests include two different distributions of the matter
\begin{itemize}
\item Spherical shell that falls radially into the black hole,
\item Localized blob of matter falling radially towards the black hole.
\end{itemize}
The tests are performed for two different equations of state
\begin{itemize}
\item Dust ($p=0$)
\item Fluid with $p \propto \rho^{1.4}$.
\end{itemize}

\subsection{Initial and boundary data for the matter}

We assume 
\begin{equation}
   \rho(u=0,r,x^A) = \left\{ \begin{array}{ll}
  \displaystyle{\lambda \exp{\left(- \frac{R_b - R_a}{2(r-R_a)} \right)} \,
                        \exp{\left(- \frac{R_b - R_a}{2(R_b-r)} \right)} \;
                        G(x^A) }
                    & \mbox{if $r  \in [R_a,R_b]$} \\
                    & \\
                  0 & \mbox{otherwise,}
                        \end{array}
                        \right.
\end{equation}
If $G(x^A)=1$, $\rho$ describes a spherical shell of matter between $r=R_a$ and 
$r=R_b$, and centered about $r=0$. We also consider the case with $G$ defined as 
a localized gaussian-type function
\begin{equation}
   G(x^A) = \left\{ \begin{array}{ll}
  \displaystyle{ ( q^2 + p^2 - \mu )^4 }
                    & \mbox{if $q^2 + p^2  \le \mu$} \\
                    & \\
                  0 & \mbox{otherwise,}
                        \end{array}
                        \right. 
\label{eq:GxA}
\end{equation}
which we use to describe an initial ``blob'' of matter. To provide initial data 
for the velocity field, we set
$V_{ang}=0$ and 
\begin{equation}
v_{rn}(u=0,r,x^A) =  -(\sqrt{1+E} + \sqrt{E+2/r}) (1-2/r)
\end{equation}
where $v_{rn}$ is $v_1$ renormalized to be well behaved on the event horizon 
$r=2$ (explicitly $v_{rn}=v_1 (1-2/r)^2$). $E$ represents the energy at infinity 
of a unit mass particle, in the sense that at infinity $|v^1|^2=E$; note that 
$E>-1$. In the tests described below we take
\begin{itemize}
\item $R_a=4$ and $R_b=7$
\item $\lambda$ varying between $10^{-5}$ and $10^{-13}$
\item $G(x^A)=1$ (spherical shell), or $G(x^A)$ given by Eq. (\ref{eq:GxA}) with 
$\mu=0.4$ (blob of matter with center at one of the poles whose density goes to 
zero at about $\theta = 44^o$)
\item $E=2.25$
\end{itemize} 

Analytically, the matter never reaches $r=2$, but in practice, due to numerical 
diffusion, boundary conditions are still needed there; we impose
\begin{equation}
\rho=0, \; \; V_{ang}=0, \; \; v_{rn}=0.
\end{equation}

\subsection{Equation of state}

For dust, the equation of state is $p=0$. We also investigate how the code copes 
with non zero pressure, and set up a test case with the density defined as above 
(with $\lambda = 10^{-9}$), and with the equation of state
\begin{equation}
p = 10^{-11} \rho^{1.4}.
\end{equation}
Further, in order to keep $p_\rho$ (Eq. (\ref{eq:prho})) well-behaved for
small $\rho$, we set $p=0$ whenever $\rho < \rho_{max} 10^{-5}$, where
$\rho_{max}$ is the maximum value of $\rho$ at $u=0$.
Although the magnitude of this pressure is rather small, it is encouraging to
see that the code can handle it even though our implementation of the
fluid equations is rather simple. More realistic equations of state would
induce shocks that would recquire a sophisticated treatment of the matter
equations. We defer this to future work.

\subsection{Physical interpretation of the initial data}

First, it is necessary to clarify the issue of units. We are using geometric 
units in which $G=c=1$, and everything is given in terms of a unit of length. 
However, this unit is {\em not} metres, but rather it is the distance 
corresponding to a unit change in the radial coordinate $r$. We will call the 
geometric unit of length $1L$, and in order to convert to S.I. units we need to 
know the value of $1L$ in metres. We write
\begin{equation}
1L=r_0 \; \; \mathrm{[metres].}
\end{equation}
For example, if we have a $10 M_\odot$ black hole with event horizon at $r=2$, 
then $r_0$ would be 14766 [metres]. We use the notation that quantities in 
geometric 
units will be denoted without suffix, and those in S.I. units will have the 
suffix ${}_S$. Then the conversions for speed $v$, mass $M$, density $\rho$
and pressure $p$ are:
\begin{eqnarray}
v \times 2.998 \times 10^8 & &= v_S \; \; \mathrm{[m/s]} \\
M \times 1.347 r_0 \times 10^{27} & &= M_S \; \; \mathrm{[kg]} \\
\rho \times \frac{1.347 \times 10^{27}}{r_0^2} & &= \rho_S \; \;
\mathrm{[kg/m^3]} \\
p \times \frac{1.210 \times 10^{44}}{r_0^2} & &= p_S \; \; \mathrm{[Pa]}
\end{eqnarray}
Note that $1.347 \times 10^{27}$ is $c^2/G$, and $1.210 \times 10^{44}$ is
$c^4/G$ in S.I. units.

We now use these conversions to determine, in S.I. units, the various parameters 
describing the initial data specified above, when the mass of the black hole is 
$10M_\odot$. For the case $\lambda = 10^{-9}$ we find
\begin{itemize}

\item {\em Spherical shell:} $\rho_S= 8.34 \times 10^8 kg/m^3$,
$v = 0.90c$, $p_S=0$ or $p_S =8.47 \times 10^{10} Pa$,
$M=3.37 \times 10^{-6} M_\odot$;
\item {\em Localized pulse:} $\rho_S= 2.14 \times 10^7 kg/m^3$,
$v = 0.90c$, $p_S=0$ or $p_S =5.01 \times 10^{8} Pa$,
$M=6.11 \times 10^{-9} M_\odot$;
\end{itemize}
where $v$ is the proper inward radial velocity, and the values of $\rho_S$,
$p_S$ and $v$ are given at the point of maximum density at $u=0$. For
$\lambda\neq 10^{-9}$, $\rho_S$ and $M$ scale linearly in $\lambda$; $p_S$
scales as $\lambda^{1.4}$; and $v$ is unchanged.

\subsection{Spherical collapse}
\begin{figure}
\centerline{\epsfxsize=6in\epsfbox{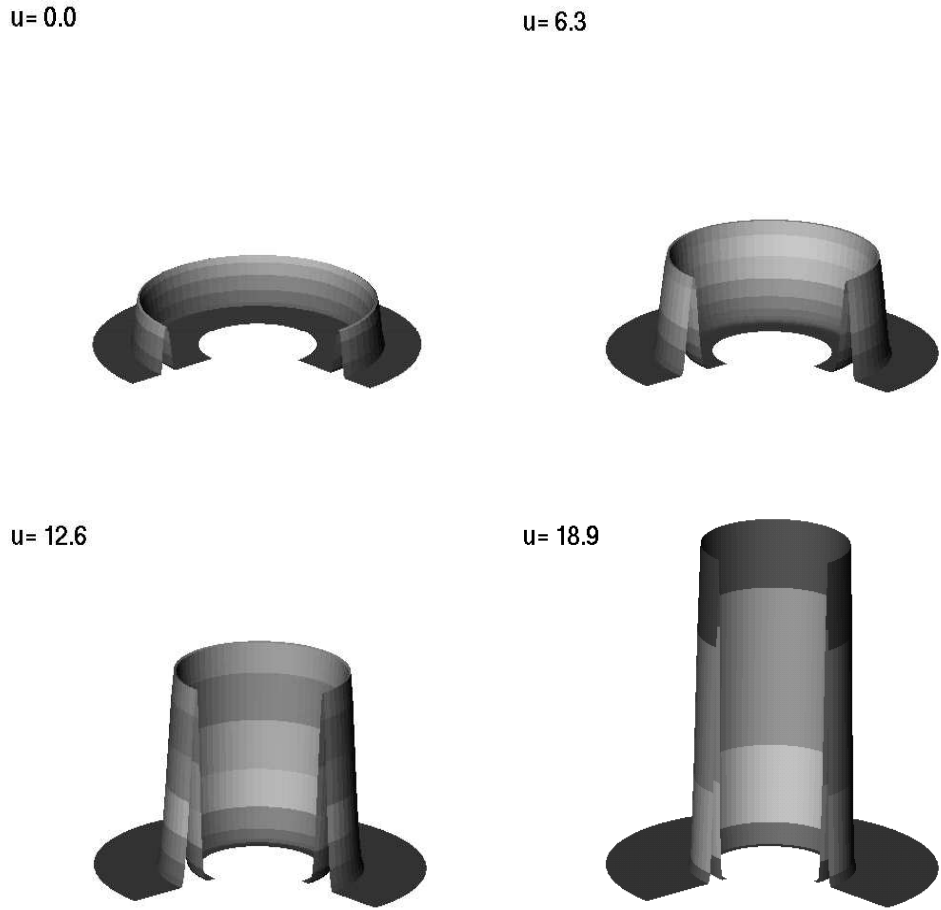}}
\caption{Density profiles vs. time. The figure shows $4$ different
snapshots of $\rho$ at the equator. The inner part ``hole'' in the
pictures corresponds to the black hole $r=2m$ radius while the outer
part corresponds to $r=\infty$. The collapse of matter onto the
black hole can be clearly seen. The plots are for the case $\lambda=10^{-9}$ and 
$p \neq 0$.}
\label{fig:rho_sph}
\end{figure}

The first test consisted in the collapse of a spherical shell onto a black hole
We set the black hole mass $=1$, the inner shell $R_a=4$ and the outer 
shell $R_b=7$. Since the resulting spacetime is spherically symmetric there 
should
be no gravitational waves emitted by the system. We confirmed this behavior
by monitoring the News function for different initial amplitudes $\lambda= 
10^{-(2n+5)}$  (with $n=0..4$)  over time. The value of each polarization
mode converges to zero as the discetrization gets refined. The matter collapses
onto the black hole and the run proceeds smoothly. Figure \ref{fig:rho_sph} 
shows
the evolution of $\rho$ for the case with pressure.

In this case it is fairly straightforward to calculate analytically the Bondi 
mass of the system, and using Maple we found that (for $p=0$, $\lambda=10^{-6}$) 
$M=1.0007395$. By increasing the resolution, we checked that this value is
approached, with second order accuracy, by the value computed with the code.
For instance, with a grid of $N_x = 165$ and $N_{\xi}=65$, the obtained value for
the mass is $M=1.0007408$; which agrees quite well with the expected value.

\subsection{Black hole-matter ball collision}
\begin{figure}
\centerline{\epsfxsize=6in\epsfbox{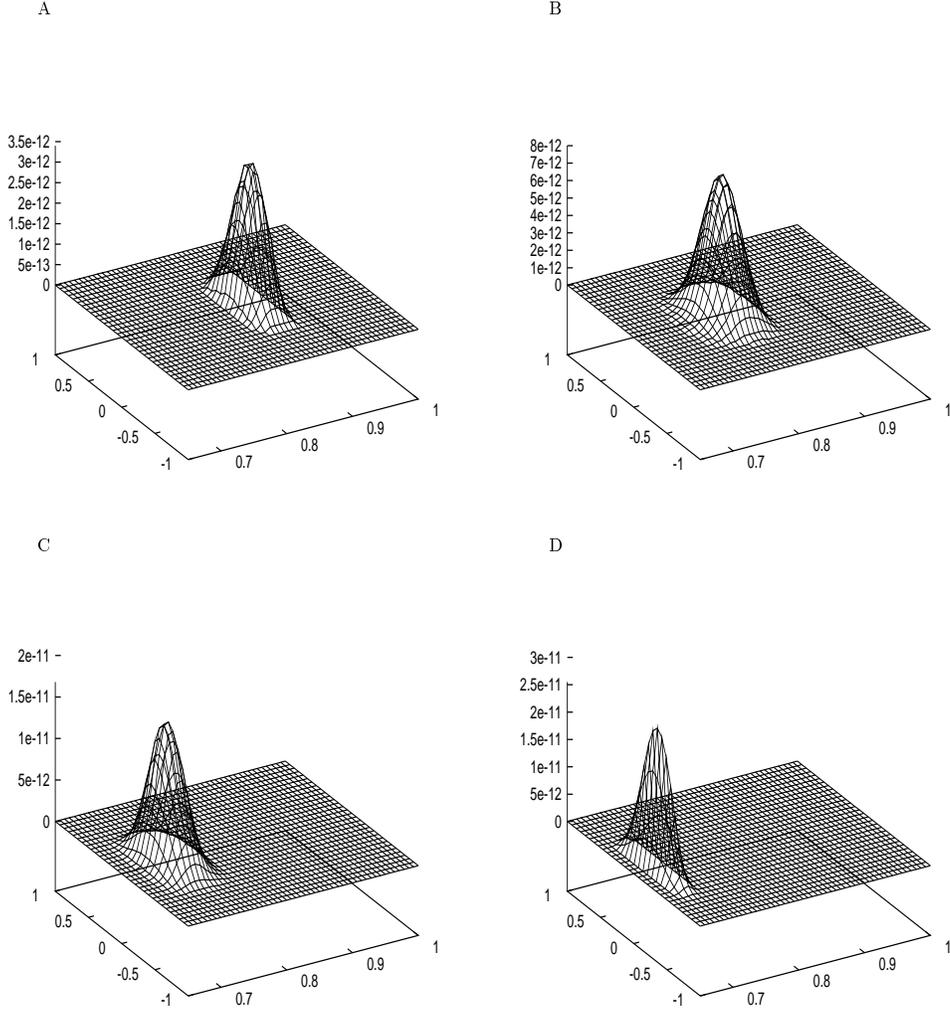}}
\caption{Density profiles vs. time for the localized ``blob''
of matter at times $u=0$ (A), $u=7.2$ (B), $u=14.3$ (C) and
$u=21.5$ (D). The value of initial amplitude parameter
for this case is $\lambda=10^{-9}$, and $p \neq 0$. The figure shows $4$ 
different snapshots
of $\rho$ at $y=0$ (on the northern hemisphere) as a function of the 
compactified
radial coordinate ($x=2/3$
being the black hole radius and $x=1$ corresponding to null infinity). The
pulse collapses onto the black hole remaining localized.}
\label{fig:rho_nosph}
\end{figure}

\begin{figure}
\centerline{\epsfxsize=6in\epsfbox{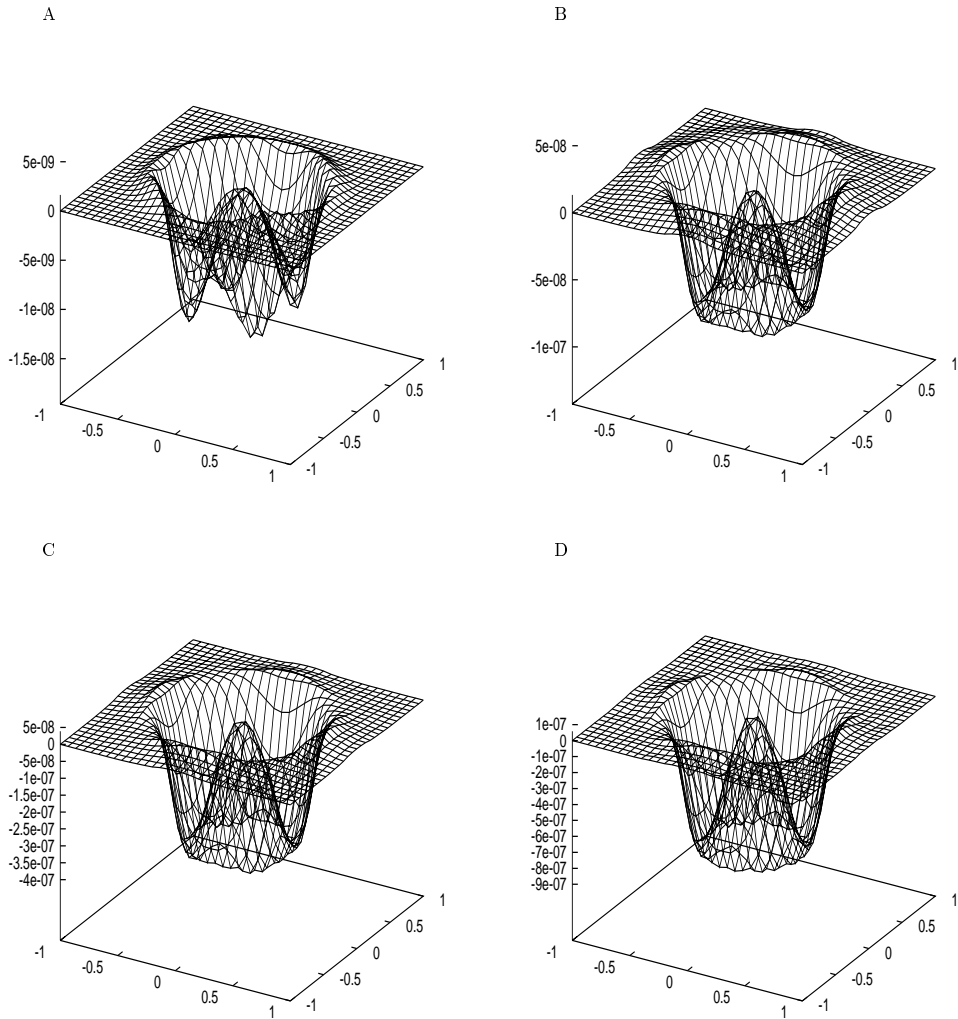}}
\caption{The ``plus'' component of the news on the northern hemisphere for
the case $\lambda=10^{-9}$ and $p \neq 0$.at times $u=0$ (A), $u=7.2$ (B), 
$u=14.3$ (C) and $u=21.5$ (D).}
\label{fig:news}
\end{figure}

To study the collision of a dust ball-black hole system we again
set the black hole mass $M=1$ and the inner and outer radius
of the blob with $R_a=4$ and $R_b=7$; finally, the localization on the
sphere is set by choosing $\mu = 0.4$. This configuration describes
a ball of dust with center at one of the poles that goes to zero at about
$\theta = 44^o$ with compact support in $[4,7]$ in the radial direction.
Note that in this case, although we initially set $V_{ang}=0$, the self
gravitational field of the dust ball induces a non zero $V_{ang}$ towards
the pole. Figure \ref{fig:rho_nosph} displays the evolution of $\rho$ for
the case with pressure ($\lambda=10^{-9}$). Again, the evolution proceeds
smoothly as the ``blob'' of matter collapses on to the black hole and the
plus polarization mode of the news is obtained at null infinity (see figure
\ref{fig:news}).

\begin{figure}
\centerline{\epsfxsize=6in\epsfbox{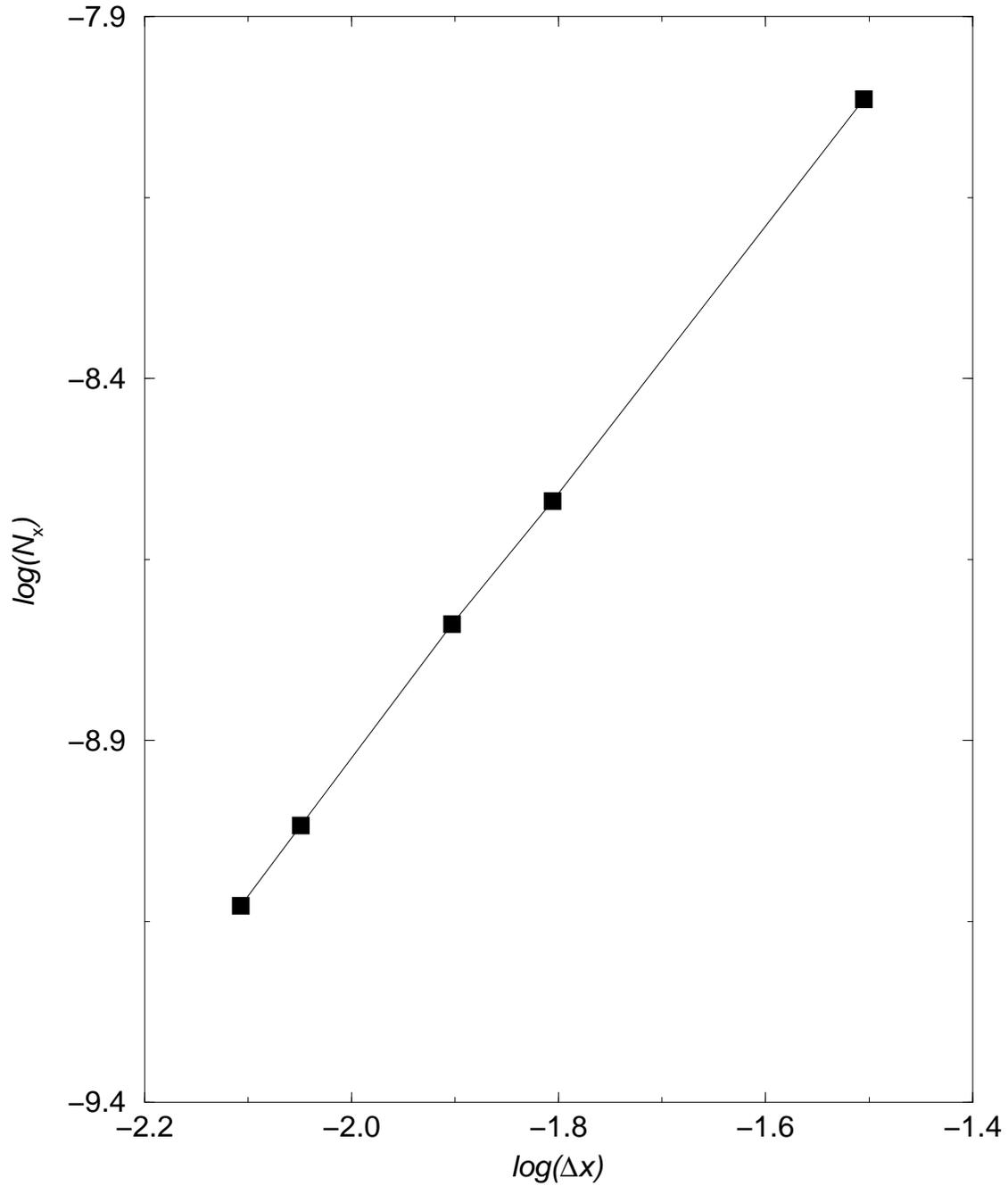}}
\caption{Convergence of the {\it cross} polarization mode in
an axysimmetric spacetime. The figure shows the logarithm of
$N_{\times}(u=0.15)$ vs. the logarithm of the discretization size. The slope
of $1.9$ is in good agreement with second order convergence. The plot is for the 
case $\lambda=10^{-9}$ and $p \neq 0$.}
\label{fig:conv}
\end{figure}

Although an analytic solution to this problem is unknown, one can check
consistency of the obtained simulation by observing that the {\em cross} 
polarization
mode (${\cal N}_{\times}$) must vanish (since the problem is still 
axysimmetric).
We confirmed this behavior, by increasing the number of grid points
and plotting the logarithm of the $L_{\infty}$ norm of ${\cal N}_{\times}$ (at 
$u=0.15$)
vs. discretization size and observing its convergence to zero. As shown in
figure \ref{fig:conv} the slope of $1.9$ is in agreement with second order
convergence. The convergence test was performed with $\lambda=10^{-9}$ and 
with non-zero pressure. We should add, however, that at later times ($u \ge 
10$), the convergence rate is reduce to about $1.5$. This is expected because of 
the non-centered scheme used at the black hole boundary (see section 
\ref{s:num}).

Another check made was that the path of the peak density should be a geodesic of 
the background spacetime. This was indeed found to be the case. For example, 
using Maple we found that at $u=2$ $r$ should be $4.9169$; and numerically we 
found, for the case $\lambda=10^{-7}$ and non-zero pressure, that $r=4.9180$ (for
a grid with $N_x=85$ and $N_{\xi}=65$).

\section{Conclusion}
\label{s:conc}

In this paper, we have incorporated matter, in the form of a perfect fluid, into 
the characteristic code for the Einstein field equations: we have found explicit 
forms for the various evolution and field equations, we have shown how these 
equations are discretized, and we have carried out a number of tests on the 
resulting code. The code is stable and convergent, and its validation 
has included the following tests
\begin{itemize}
\item The peak of the matter ``blob'' follows a geodesic of the background 
spacetime.
\item The code is not written with any symmetries, yet symmetric initial data 
leads to the appropriate part of the gravitational radiation vanishing.
\item The initial mass of the system, as calculated by the code, agrees with the 
initial mass calculated analytically.
\end{itemize}

This paper has not computed a solution to any real problem in astrophysics. 
Nevertheless, we have shown that the present code is capable of modeling a 
variety of situations where matter is captured by a black hole; although our 
treatment of the hydrodynamics would need to be amended in order to be able to 
handle shock waves. Even so, the present code should be able to contribute 
towards an understanding of a number of problems in black hole astrophysics.

\section*{ACKNOWLEDGEMENTS}

The authors thank Philippos Papadopoulos, Jose Font, Richard Matzner, Matthew
Choptuik and Ed Seidel for helpful comments. This work has been supported by NSF PHY
9510895 and NSF INT 9515257 to the University of Pittsburgh, by NSF PHY 9800722 and
NSF PHY 9800725 to the University of Texas at Austin and by the Binary Black Hole
Grand Challenge Alliance, NSF PHY/ASC 9318152. Computer time  was provided by the
Pittsburgh Supercomputing Center and the San Diego Supercomputer Center. L.L. was
partially supported by a Mellon Predoctoral Fellowship at the  University of
Pittsburgh. N.T.B. and M.M. thank the Foundation  for Research Development, South
Africa, for financial support. N.T.B. thanks the  University of Pittsburgh for
hospitality. L.L. thanks the Universities of South  Africa and of Durban-Westville
for their hospitality.

\section*{APPENDIX}

The evolution equations for the matter variables are given below. Note that
$V_w$ is $1+W/r$.
\begin{eqnarray}
V_{ang,u} & = &
\frac{p_\rho}{4 v_1 r^2 } \bigg( (2 e^{2\beta} (\bar{\eth} \rho) V_{ang}^2
 +2 e^{2\beta} V_{ang} (\eth \rho) \bar{V}_{ang} ) K
 -4 V_{ang} \rho_{,r} v_0 r^2
 -2 V_{ang} \rho_{,r} \bar{V}_{ang} U r^2 \nonumber \\
& - & 4 V_{ang} \rho_{,u} v_1 r^2
 +4 e^{2\beta} (\eth \rho) r^2
 -2 e^{2\beta} (\eth \rho) V_{ang}^2 \bar{J}
 -2 V_{ang} v_1 (\eth \rho) \bar{U} r^2
 -2 V_{ang} v_1 (\bar{\eth} \rho) U r^2 \nonumber \\
& + & 4 V_{ang} \rho_{,r} v_1 V_w r^2
 -2 V_{ang}^2 \rho_{,r} \bar{U} r^2
 -2 e^{2\beta} V_{ang} (\bar{\eth} \rho) \bar{V}_{ang} J
\bigg) \nonumber \\
& + & \frac{1}{4 v_1 r^2 } \bigg(
  4 v_1 r^2 U (\eth \beta) \bar{V}_{ang}
 -e^{2\beta} (\eth J) \bar{V}_{ang}^2
 -e^{2\beta} (\eth \bar{J}) V_{ang}^2
 +4 v_1 r^2 \bar{U} (\eth \beta) V_{ang} \nonumber \\
& -&4 V_{ang,r} v_0 r^2
 +  4 V_w V_{ang,r} v_1 r^2
 -2 e^{2\beta} (\eth V_{ang}) V_{ang} \bar{J}
 -2 v_1 r^2 (\eth \bar{U}) V_{ang} \nonumber \\
& + & (
 2 e^{2\beta} (\eth V_{ang}) \bar{V}_{ang}
 +2 e^{2\beta} (\bar{\eth} V_{ang}) V_{ang}
 ) K
 -4 V_w v_1^2 r^2 (\eth \beta)
 -2 e^{2\beta} (\bar{\eth} V_{ang}) \bar{V}_{ang} J \nonumber \\
& + & 2 e^{2\beta} J K (\eth \bar{J}) V_{ang} \bar{V}_{ang}
 +2 e^{2\beta} \bar{J} K (\eth J) V_{ang} \bar{V}_{ang}
 -2 v_1 U (\bar{\eth} V_{ang}) r^2 \nonumber \\
& -&2 e^{2\beta} (\eth K) V_{ang} \bar{V}_{ang}
 + 2 v_1^2 r^2 (\eth V_w)
 -4 e^{2\beta} \bar{J} J (\eth K) V_{ang} \bar{V}_{ang}
 -2 V_{ang,r} U \bar{V}_{ang} r^2 \nonumber \\
& +&8 v_0 v_1 r^2 (\eth \beta)
 -2 v_1 \bar{U} (\eth V_{ang}) r^2
 - 2 r^2 v_1 (\eth U) \bar{V}_{ang}
 -2 V_{ang,r} \bar{U} V_{ang} r^2 \bigg)
\end{eqnarray}
\begin{equation}
\rho_{,u}=-\frac{e^{2\beta}(F_o v_1 -F_a (p + \rho))}
{v_1^2(p_\rho (p + \rho) -1)}, \; \; \;
v_{1,u}=\frac{ e^{2\beta}( -F_a +F_o v_1 p_\rho)}{v_1(p_\rho (p + \rho) -1)}
\end{equation}
with $F_o$ and $F_a$ given by
\begin{eqnarray}
F_o &=&
\frac{p + \rho}{4 e^{2\beta} r^2} \bigg(
(
 -2 (\eth K) \bar{V}_{ang} e^{2\beta}
 -2 (\bar{\eth} K) V_{ang} e^{2\beta}
 ) J \bar{J}
 -2 (\eth \bar{J}) V_{ang} e^{2\beta} \nonumber \\
& +&(
 (\eth J) \bar{V}_{ang} e^{2\beta}
 +(\bar{\eth} J) V_{ang} e^{2\beta}
 ) K \bar{J}
 +(
 (\eth \bar{J}) \bar{V}_{ang} e^{2\beta}
 +(\bar{\eth} \bar{J}) V_{ang} e^{2\beta}
 ) K J \nonumber \\
& +&8 v_1 r V_w
 -4 v_{0,r} r^2
 -2 v_1 r^2 (\bar{\eth} U)
 +(
 -2 (\eth V_{ang}) e^{2\beta}
 -4 (\eth \beta) V_{ang} e^{2\beta}
 ) \bar{J}
 +4 V_w v_{1,r} r^2 \nonumber \\
& -&2 (\bar{\eth} J) \bar{V}_{ang} e^{2\beta}
 +(
 -4 (\bar{\eth} \beta) \bar{V}_{ang} e^{2\beta}
 -2 (\bar{\eth} \bar{V}_{ang}) e^{2\beta}
 ) J \nonumber \\
& +&(
 2 (\eth \bar{V}_{ang}) e^{2\beta}
 +4 (\bar{\eth} \beta) V_{ang} e^{2\beta}
 +4 (\eth \beta) \bar{V}_{ang} e^{2\beta}
 +2 (\bar{\eth} V_{ang}) e^{2\beta}
 ) K \nonumber \\
& -&2 U \bar{V}_{ang,r} r^2
 -2 \bar{U} V_{ang,r} r^2
 +4 v_1 r^2 V_{w,r}
 -8 v_0 r
 -2 r^2 \bar{U}_{,r} V_{ang}
 -2 U (\bar{\eth} v_1) r^2 \nonumber \\
& -&4 r \bar{U} V_{ang}
 -4 r U \bar{V}_{ang}
 -2 r^2 U_{,r} \bar{V}_{ang}
 -2 v_1 r^2 (\eth \bar{U})
 -2 \bar{U} (\eth v_1) r^2 \bigg) \nonumber \\
& +&\frac{1}{4 e^{2\beta} r^2} \bigg(
 -2 \rho_{,r} V_{ang} \bar{U} r^2
 -2 (\bar{\eth} \rho) \bar{V}_{ang} e^{2\beta} J
 -4 \rho_{,r} v_0 r^2
 -2 (\eth \rho) V_{ang} e^{2\beta} \bar{J}
 -2 \rho_{,r} \bar{V}_{ang} U r^2 \nonumber \\
& -&2 v_1 (\bar{\eth} \rho) U r^2
 +(
 2 (\bar{\eth} \rho) V_{ang} e^{2\beta}
 +2 (\eth \rho) \bar{V}_{ang} e^{2\beta}
 ) K
 +4 \rho_{,r} v_1 V_w r^2
 -2 v_1 (\eth \rho) \bar{U} r^2 \bigg)
\end{eqnarray}
\begin{eqnarray}
F_a& =&\frac{p_\rho}{4 e^{2\beta} r^3} \bigg(
 -2 v_1 \rho_{,r} \bar{V}_{ang} U r^3
 -2 v_1^2 (\eth \rho) \bar{U} r^3
 -2 v_1^2 (\bar{\eth} \rho) U r^3
 -4 v_1 \rho_{,r} v_0 r^3
 +4 v_1^2 \rho_{,r} V_w r^3 \nonumber \\
& -&2 r v_1 (\bar{\eth} \rho) \bar{V}_{ang} e^{2\beta} J
 -2 r v_1 (\eth \rho) V_{ang} e^{2\beta} \bar{J}
 +(
 2 r v_1 (\bar{\eth} \rho) V_{ang} e^{2\beta}
 +2 r v_1 (\eth \rho) \bar{V}_{ang} e^{2\beta}
 ) K \nonumber \\
& -&2 v_1 \rho_{,r} V_{ang} \bar{U} r^3
 +4 \rho_{,r} r^3 e^{2\beta} \bigg) \nonumber \\
& +&\frac{1}{4 e^{2\beta} r^3} \bigg(
 (
 -2 r (\eth v_1) V_{ang} e^{2\beta}
 +2 V_{ang}^2 e^{2\beta}
 ) \bar{J}
 +8 \beta_{,r} v_1 v_0 r^3
 -2 v_1 U (\bar{\eth} v_1) r^3
 +2 v_1^2 r^3 V_{w,r} \nonumber \\
& -&2 v_{1,r} \bar{U} V_{ang} r^3
 +4 \beta_{,r} v_1 r^3 \bar{U} V_{ang}
 -2 v_1 r^3 \bar{U}_{,r} V_{ang}
 +4 V_w v_{1,r} v_1 r^3
 -2 v_{1,r} U \bar{V}_{ang} r^3 \nonumber \\
& +&K J r \bar{J}_{,r} V_{ang} \bar{V}_{ang} e^{2\beta}
 -2 v_1 \bar{U} (\eth v_1) r^3
 +4 \beta_{,r} v_1 r^3 U \bar{V}_{ang}
 -4 v_{1,r} v_0 r^3
 -r \bar{J}_{,r} V_{ang}^2 e^{2\beta} \nonumber \\
& +&(
 -4 V_{ang} \bar{V}_{ang} e^{2\beta}
 +2 r (\eth v_1) \bar{V}_{ang} e^{2\beta}
 +2 r (\bar{\eth} v_1) V_{ang} e^{2\beta}
 ) K
 -4 V_w \beta_{,r} v_1^2 r^3 \nonumber \\
& -&2 v_1 r^3 U_{,r} \bar{V}_{ang}
 +(
 -2 r (\bar{\eth} v_1) \bar{V}_{ang} e^{2\beta}
 +2 \bar{V}_{ang}^2 e^{2\beta}
 ) J
 -2 J \bar{J} r K_{,r} V_{ang} \bar{V}_{ang} e^{2\beta} \nonumber \\
& +&K \bar{J} r J_{,r} V_{ang} \bar{V}_{ang} e^{2\beta}
 -r J_{,r} \bar{V}_{ang}^2 e^{2\beta} \bigg).
\end{eqnarray}

\end{document}